\documentclass[twocolumn,showpacs,preprintnumbers,amsmath,amssymb]{revtex4}
\usepackage{graphicx}
\usepackage{dcolumn}
\usepackage{bm}
\usepackage{multirow}

\begin{document}

\preprint{}

\title{Hybrid Monte Carlo/molecular dynamics simulation of a refractory metal high entropy alloy}

\author{M. Widom}
\author{W. P. Huhn}
\affiliation{
Department of Physics, Carnegie Mellon University,
Pittsburgh, PA  15213
}
\author{S. Maiti}
\author{W. Steurer}
\affiliation{
Laboratory of Crystallography, 
ETH Zurich,
CH-8093 Zurich, Switzerland
}

\date{\today}
\begin{abstract}
The high entropy alloy containing refractory metals Mo-Nb-Ta-W has a body centered cubic structure, which is not surprising given the complete mutual solubility in BCC solid solutions of all pairs of the constituent elements.  However, first principles total energy calculations for the binaries reveal a set of distinct energy minimizing structures implying the likelihood of chemically ordered low temperature phases.  We apply a hybrid Monte Carlo and molecular dynamics method to evaluate the temperature-dependent chemical order.  Monte Carlo species swaps allow equilibration of the structure that cannot be achieved by conventional molecular dynamics.  At 300~K a cesium-chloride ordering emerges between mixed (Nb,Ta) sites and mixed (Mo,W) sites.  This order is lost at elevated temperatures.
\end{abstract}

\maketitle

\section{Introduction}
High entropy alloys~\cite{Yeh04} are multicomponent mixtures of elements capable of mutual chemical substitution, allowing in principle a configurational entropy of $\log{N}$ per atom for an equiatomic $N$-component system. This entropy is believed responsible for stabilizing simple Bravais lattice-type sructures at high temperatures, in which every atomic site is equivalent.  A typical example is provided by the refractory element compound Mo-Nb-Ta-W~\cite{Senkov10}.  All four constituents individually take the body-centered cubic structure (Strukturbericht A2, Pearson type cI2), and their six binary combinations form complete BCC solid solutions.  If every lattice site were to be occupied with equal probability by each of the four elements, a configurational entropy of $k_B\ln{4}\sim 1.2\times 10^{-4}$~eV/K would result.  However, atomic size, enthalpy of mixing and additional considerations surely affect stability also~\cite{Guo2011,Yang2012,Zhang2012,Otto2013,Tian2013}.  Indeed, decomposition into multiple phases has been observed experimentally in certain high entropy alloy families~\cite{Singh2011,Chen2006,Zhang2013}.

Competing against this entropy is the energetic preference for chemical order among the constituent elements.  Although no chemically ordered solid phases are known, first principles total energy calculations reveal energetic preferences for alternation of chemical occupancy on neighboring sites, leading, for example, to the cesium-chloride structure (Strukturbericht B2, Pearson type cP2) in which one class of atoms prefers to occupy cube vertices and another class of atoms prefers cube centers.  Given this competition it is probable that an order-disorder transition would occur at finite temperature.  The fact that ordered states have not been observed probably indicates that this transition temperature is sufficiently low that diffusion is inhibited and the system has fallen out of equilibrium.  However, vestiges of the preferred order may remain even at elevated temperatures in the form of short-range correlations that reduce the entropy below its theoretical maximum.

In order to probe the chemical ordering we perform hybrid Monte Carlo/molecular dynamics (MC/MD) simulations to explore the evolution of chemical order as a function of temperature.  Our results show that atomic size differences (e.g. Nb and Ta are larger than Mo and W) favor chemical ordering at low temperature, and a surprising degree of disorder among atoms of similar size (e.g. Nb with Ta, and Mo with W) persists even below T=300~K.  As temperature rises, up to a maximum temperature investigated of T=1800~K, vestiges of chemical ordering remain and the configurational entropy remains bounded below the theoretical maximum.

\section{Chemical species and their pairwise interactions}

The four elements belong to a $2\times 2$ square block of the periodic table, and hence their physical properties will bear specific relationships to each other.  The present study considers elements from the chromium group (group 5, here Ta and Nb) and from the vanadium group (group 6, here W and Mo) located in the fourth row (here Nb and Mo) and in the fifth row (here Ta and W).  Atomic size drops as group number increases, and to a lesser as row number increases.  Hence, we understand the sequence of atomic volumes (in units of~\AA$^3$/atom): Ta (18.10) $\gtrsim$ Nb (18.05) $\gg$ W (15.82) $\gtrsim$ Mo (15.61).  Trends in electronegativity (Pauling scale) are nearly (but not perfectly) opposite to atomic volume, hence: Ta (1.5) $\lesssim$ Nb (1.6) $\ll$ Mo (2.16) $\lesssim$ W (2.36).  Various combinations of size and electronegativity differences have been linked to structure formation, and the present compounds lie close to a predicted cP2-forming region~\cite{Villars83,Pettifor88}.  In this paper we will mainly refer to atomic size, but in fact both size and electronegativity play mutually reinforcing roles.

Owing to these volume and electronegativity contrasts we anticipate relatively strong binding between elements from different transition metal groups of the periodic table (here Nb-Mo, Nb-W, Ta-Mo and Ta-W), and relatively  weak binding between elements from the same group (Nb-Ta and Mo-W).   Among the inter-group formation enthalpies, we expect the pair Nb-W, with intermediate sizes and electronegativities, to be the weakest interaction, and this expectation is reflected in the strong negative deviation of the Nb-W solidus temperature relative to Vegard's rule.

An exhaustive ground state search has been performed for the strongly interacting binaries drawn from different groups using the LDA functional to generate a cluster expansion for interatomic interactions~\cite{Blum05}.  Other researchers have examined specific compounds using similar methods~\cite{Curtarolo08,Turchi01,Turchi04,Hart05}.  We have repeated this search and supplemented it with an investigation of the weakly interacting binaries drawn from the same groups.  The results are summarized in Table~\ref{tab:pairs}, based on the VASP~\cite{Kresse96} calculations using projector augmented wave potentials~\cite{Kresse99} with the PBE~\cite{Perdew96} exchange correlation functional.  The enthalpies of formation increase away from the diagonal in this table, reflecting the greater tendency towards compound formation associated with greater atomic size contrast.  The most common energy minimizing structure is Pearson type oA12 (Strukturbericht B2$_3$) which is cP2-like but with antiphase boundaries.  On average the perfectly ordered cP2 structure is about 8 meV/atom (i.e. $k_B$T at about T=100~K) above the energy minimizing structure.

\begin{table}
\begin{tabular}{r|ll|ll}
   &\multicolumn{2}{c|}{Group 5}&\multicolumn{2}{c}{Group 6}\\
   &  Ta        & Nb        & W         & Mo \\
\hline
Ta &  cI2 (0)   &  hR8 (4)   & oA12 (103) & oA12 (186) \\
Nb &  hR8 (4)   &  cI2 (0)   & cF16 (47)  & oA12 (103) \\
\hline
W  & oA12 (103) & cF16 (47)  &  cI2 (0)   &  cP2 (9)   \\
Mo & oA12 (186) & oA12 (103) &  cP2 (9)   &  cI2 (0)   \\
\end{tabular}
\caption{\label{tab:pairs} Optimal structures and enthalpies for pairwise interactions at 50-50\% composition. Elements are ordered in decreasing atomic volume.  Pearson symbol of stable compound followed by negative enthalpy of formation (units meV/atom).}
\end{table}

Consequently, in our four element compound, we anticipate low energy structures will exhibit chemical order dominated by cP2-like alternation of (Nb,Ta) atoms on one sublattice (e.g. cube vertices) and (Mo,W) atoms on the other sublattice (e.g. cube body centers).  The absolute ground state may in fact separate into binaries or ternaries of various compositions, but owing to the strong interactions across groups, and the weak interactions within groups, we expect to achieve cP2-like order even at relatively low temperatures.

\section{Hybrid Monte Carlo/molecular dynamics}

Molecular dynamics is well suited to reproduce the small amplitude oscillations of atoms in the vicinity of crystal lattice sites.  At low temperatures the probability for an atom crossing the barrier from one lattice site to another is prohibitively low and will rarely occur on the time scale of a molecular dynamics run.  In contrast, Monte Carlo swaps of atomic species on different sites occurs with a probability $P=\exp{(-\Delta E/k_BT)}$ related to the net energy difference $\Delta E=E_{swap}-E_{ini}$ of swapped and initial configurations.  In particular, it is independent of the energy barrier separating the states.  Since some of the pairwise interactions presented in Table~\ref{tab:pairs} are quite low, we expect that some Monte Carlo species swaps will be accepted even at temperatures below T=300~K.

We implement this computationally by alternating molecular dynamics with Monte Carlo swaps, each performed from first principles using VASP. In the runs described below we perform 10 MD steps, with a 1 fs time step, between each attempted species swap.  We prepare an initial configurations starting with a 16-atom $2\times 2\times 2$ BCC supercell with a random distribution of atomic species which we thoroughly anneal under MC/MD at a given temperature.  We use annealed structures to create successively larger supercells, of size 32, 64 and eventually 128-atoms, thoroughly annealing at each size.  Following the annealing, run durations for data collection are 10~ps, and include 1000 attempted species swaps. The volume is held at 16.9~\AA$^3$/atom throughout, a value obtained consistently when fully relaxing samples equilibrated at T=300~K.

\begin{figure}
\includegraphics[width=2.5in,angle=-90]{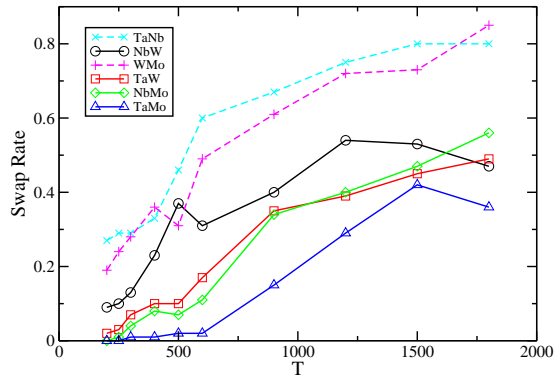}
\vspace{1cm}
\caption{\label{fig:rates} Monte Carlo acceptance rates (ratio of accepted to attempted species swaps) for each species pair vs. temperature.}
\end{figure}

The feasibility of the MC/MD method depends on adequate acceptance rates for attempted species swaps.  As illustrated in Fig.~\ref{fig:rates}, it is evident that many swaps are accepted with reasonably high probability at all temperatures studied (200~K and above).  These high acceptance rates indicate that the Monte Carlo achieves our intended goal of sampling the full configurational ensemble in the solid state.  The acceleration of sampling relative to conventional molecular dynamics is effectively infinite, as the same species swap will almost never be observed via molecular dynamics at low temperatures.  The utility of various hybrid MC/MD methods has been reviewed recently~\cite{Neyts2013} although these are more often applied in conjunction with empirical interaction potentials.

Swaps with the greatest size contrast (e.g. Ta-Mo and Nb-Mo) occur most infrequently, and are nearly absent at low temperatures, while the intermediate size swaps (Nb-W) occur occasionally and the most similar size swaps (e.g. Ta-Nb and W-Mo) occur with high probability even at the lowest temperatures.  We conclude that at low temperature the system behaves nearly as a pseudobinary, consisting of the chromium group (group 5, here Ta and Nb) and the vanadium group (group 6, here W and Mo) as two effective species.

\section{Results}

\begin{figure}
\includegraphics[width=2.5in,angle=-90]{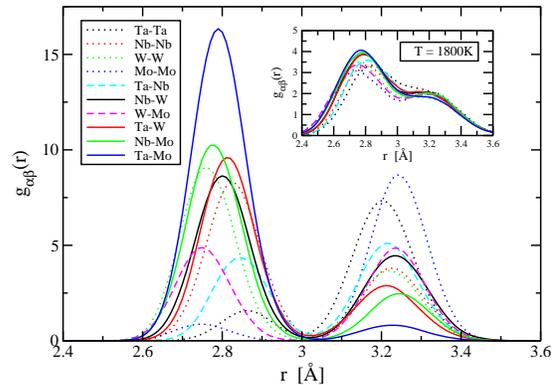}
\vspace{1cm}
\caption{\label{fig:gofr} Partial radial distribution functions $g_{\alpha\beta}(r)$ at T=300K for N=128 atoms.  Inset: T=1800K.}
\end{figure}

Given a thoroughly sampled ensemble of configurations, we can analyze a great many properties.  Of special interest are structural characterizations such as pair correlation functions and their reciprocal space equivalents, structure factors.  Fig.~\ref{fig:gofr} illustrates the partial pair correlation functions $g_{\alpha\beta}(r)$ which give the spatial distribution of atoms of species $\beta$ as a function of distance from an atom of species $\alpha$, normalized by their respective densities.  On a body centered lattice of conventional lattice constant $a$, the nearest neighbor (NN) separation $\sqrt{3}a/2$ corresponds to the cube vertex-to-body center distance, while the lattice constant $a$ itself corresponds to the next nearest neighbor (NNN) distance, vertex-to-vertex or center-to-center.  For our structures the conventional cubic lattice constant $a=3.23$~\AA~ while the near neighbor distance $\sqrt{3}a/2=2.80$~\AA.

Notice that the dominant nearest neighbor correlations are between group 5 and group 6 elements (i.e. the strongly binding Ta-W, Ta-Mo, Nb-W and Nb-Mo, solid lines in Fig.~\ref{fig:gofr}).  TaMo has the largest enthalpy for pairwise interaction (Table~\ref{tab:pairs}), and shows the strongest peak, maximizing the number of TaMo bonds.  Next come NbMo and TaW, which are nearly degenerate in enthalpy, and accordingly their corresponding correlations are similar in height, although their peak positions reflect the smaller size of fourth transition metals (Nb and Mo) relative to fifth row (Ta and W).  NbW is the weakest of the inter-group bonds and has the weakest peak.  The inter-row correlations, WMo and TaNb are both quite weak and strongly reveal the size difference between group 5 and group 6 pairs.  Among the pair correlations of like species, TaTa and MoMo are weak, as expected, however the intermediate size NbNb and WW correlations show surprising strength indicating the pseudobinary model is not perfect, suggesting the presence of some more subtle and complex order than cP2, perhaps including phase separation that cannot be fully realized in these small system sizes.

Even at T=1800~K, the highest temperature investigated, short-range chemical order is evident despite the absence of long-range order. Specifically, the inter-group near neighbor correlations remain above the inter-row correlations, which in turn remain above the self (e.g. Ta-Ta) correlations, following the patterns of interaction strengths observed in Table~\ref{tab:pairs}.

Analogous neighbor correlations were obtained previously~\cite{Grosso12} by a group using the Bozzolo, Ferrante and Smith~\cite{Bozzolo92} empirical interaction model.  Owing to their interactions, the precise details of favored and disfavored neighbor pairs differ from our present results. Interestingly, they obtain order-disorder transitions around T=600-800~K, consistent with our finding.  An order-disorder transition in Ta-W has been predicted near T=1000~K~\cite{Jindo06}.

Another popular configurational measure is the structure factor, which can actually be probed experimentally through diffraction.  This is the Fourier transform of the set of atomic positions, weighted by the scattering amplitudes of the individual atoms.  For x-ray diffraction the scattering amplitude is the form factor $f$, which may be approximated at low wavenumber as the atomic number $Z$ of the atom.  Unfortunately, the average atomic number of our group 5 atoms, $\bar{Z}$=57, hardly differs from our group 6 atoms $\bar{Z}$=58, so we expect little contrast that could reveal chemical order between groups, but stronger contrast between rows (fourth row $\bar{Z}=41.5$ compared with fifth row $\bar{Z}=73.5$).  Meanwhile, the neutron scattering lengths $b$ are quite similar for Ta, Nb and Mo, but differ for W, hence neutron diffraction is primarily sensitive to the distribution of the W atoms.  However, as is evident in the near neighbor correlation functions in Fig.~\ref{fig:gofr}, the intermediate size W (and also Nb) atoms exhibit less chemical order than the highly dissimilar Ta and Mo atoms.  To best probe the cP2-like structure it pays to define an artificial scattering amplitude, which we take as 1 for group 5 elements and 2 for group 6.

\begin{figure}
\includegraphics[width=2.5in,angle=-90]{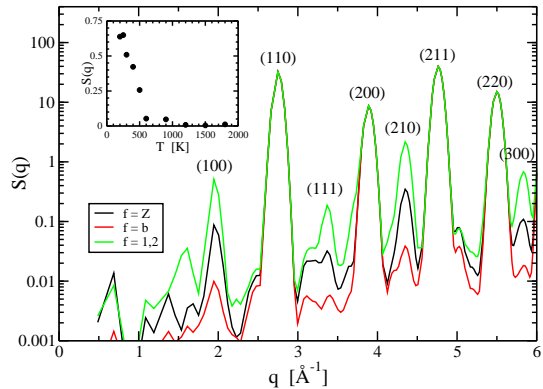}
\vspace{1cm}
\caption{\label{fig:sofq} Diffraction intensities at T=300K for N=128 atoms.  Contrast is for x-rays ($f=Z$, black), neutrons ($f=b$, green) and artificial ($f$=1 or 2 according to group 5 or 6, red). A Gaussian smearing of 0.05~\AA$^{-1}$ has been applied.  Primary reflection Miller indices sum to even integers, while odd sums correspond to superstructure peaks.  Inset, (100) peak intensity as a function of temperature for $f=1,2$.}
\end{figure}

The diffraction pattern of BCC contains only peaks whose Miller indices sum to even numbers, while cP2-like order shows up as superstructure peaks with odd sums (Fig.~\ref{fig:sofq}).  For the artificial probe that is most sensitive to the periodic table group, and to a lesser extent with the x-ray and neutron probes, we see superstructure peaks where the peak indices sum to odd numbers.  The inset shows the temperature variation of the (100) peak.  Clearly cP2-like order is present over a range of temperatures at the length scale of the studied system sizes.  Presumably the small nonmonotonicities are due to as-yet insufficient sampling.  Further studies of the scaling with system size will be needed to verify if an order-disorder is occurring.  Experimentally, diffuse (100) and (300) peaks indicating short-ranged order were observed in Ta-Mo binary alloys~\cite{Predmore70}. 

\section{Discussion}

The principal result of this work is the demonstration of the feasibility of hybrid MC/MD simulation in a first principles environment.  We have shown the method is especially well suited to the simulation of high entropy alloys, where multiple chemical species freely substitute.  Owing to the ability to swap chemical species with no barrier-related rate limitation, the method is essentially infinitely more efficient than conventional MD.  The Monte Carlo steps allow us to equilibrate chemical order in the solid state, which cannot be achieved by conventional molecular dynamics.  While the Monte Carlo dynamics is not physical, the gain in efficiency is an overwhelming advantage, provided only equilibrium properties are desired.

With respect to the application to the specific HEA, we have shown that atomic size contrast favors the development of long-ranged cP2-like chemical order between larger and smaller species.  This order is practically not visible in diffraction patterns weighted by x-ray or neutron form factors.  Even if it were feasible to observe in principle, there remains a question if the actual materials can ever reach the ordered state. We find the long-range order vanishes somewhere in the T=600-1200~K range. Improved annealing and investigation of finite size dependence will be required to identify the transition more precisely.  A reasonable estimate of the melting point is around T=3000~K, so annealing the samples at temperatures low enough that long range chemical order will develop might not be feasible.  Perhaps the best way to test these ideas would be to find other probes of short-range chemical order and study their evolution as a function of annealing temperature.

\section*{Acknowledgements}
 This work was supported in part by grant DTRA1-11-1-0064. We thank Marek Mihalkovi\v{c} and Michael Gao for useful discussions.

\bibliography{monbtaw}

\begin{thebibliography}{10}%
\makeatletter
\providecommand \@ifxundefined [1]{%
 \ifx #1\undefined \expandafter \@firstoftwo
 \else \expandafter \@secondoftwo
\fi
}%
\providecommand \@ifnum [1]{%
 \ifnum #1\expandafter \@firstoftwo
 \else \expandafter \@secondoftwo
\fi
}%
\providecommand \enquote [1]{``#1''}%
\providecommand \bibnamefont  [1]{#1}%
\providecommand \bibfnamefont [1]{#1}%
\providecommand \citenamefont [1]{#1}%
\providecommand\href[0]{\@sanitize\@href}%
\providecommand\@href[1]{\endgroup\@@startlink{#1}\endgroup\@@href}%
\providecommand\@@href[1]{#1\@@endlink}%
\providecommand \@sanitize [0]{\begingroup\catcode`\&12\catcode`\#12\relax}%
\@ifxundefined \pdfoutput {\@firstoftwo}{%
 \@ifnum{\z@=\pdfoutput}{\@firstoftwo}{\@secondoftwo}%
}{%
 \providecommand\@@startlink[1]{\leavevmode\special{html:<a href="#1">}}%
 \providecommand\@@endlink[0]{\special{html:</a>}}%
}{%
 \providecommand\@@startlink[1]{%
  \leavevmode
  \pdfstartlink
   attr{/Border[0 0 1 ]/H/I/C[0 1 1]}%
   user{/Subtype/Link/A<</Type/Action/S/URI/URI(#1)>>}%
  \relax
 }%
 \providecommand\@@endlink[0]{\pdfendlink}%
}%
\providecommand \url  [0]{\begingroup\@sanitize \@url }%
\providecommand \@url [1]{\endgroup\@href {#1}{\urlprefix}}%
\providecommand \urlprefix [0]{URL }%
\providecommand \Eprint[0]{\href }%
\@ifxundefined \urlstyle {%
  \providecommand \doi [1]{doi:\discretionary{}{}{}#1}%
}{%
  \providecommand \doi [0]{doi:\discretionary{}{}{}\begingroup
  \urlstyle{rm}\Url }%
}%
\providecommand \doibase [0]{http://dx.doi.org/}%
\providecommand \Doi[1]{\href{\doibase#1}}%
\providecommand \bibAnnote [3]{%
  \BibitemShut{#1}%
  \begin{quotation}\noindent
    \textsc{Key:}\ #2\\\textsc{Annotation:}\ #3%
  \end{quotation}%
}%
\providecommand \bibAnnoteFile [2]{%
  \IfFileExists{#2}{\bibAnnote {#1} {#2} {\input{#2}}}{}%
}%
\providecommand \typeout [0]{\immediate \write \m@ne }%
\providecommand \selectlanguage [0]{\@gobble}%
\providecommand \bibinfo [0]{\@secondoftwo}%
\providecommand \bibfield [0]{\@secondoftwo}%
\providecommand \translation [1]{[#1]}%
\providecommand \BibitemOpen[0]{}%
\providecommand \bibitemStop [0]{}%
\providecommand \bibitemNoStop [0]{.\EOS\space}%
\providecommand \EOS [0]{\spacefactor3000\relax}%
\providecommand \BibitemShut [1]{\csname bibitem#1\endcsname}%
\bibitem{Yeh04}%
  \BibitemOpen
  \bibfield{author}{%
  \bibinfo {author} {\bibfnamefont{J.~W.}\ \bibnamefont{Yeh}}, \bibinfo
  {author} {\bibfnamefont{S.~K.}\ \bibnamefont{Chen}}, \bibinfo {author}
  {\bibfnamefont{S.~J.}\ \bibnamefont{Lin}}, \bibinfo {author}
  {\bibfnamefont{J.~Y.}\ \bibnamefont{Gan}}, \bibinfo {author}
  {\bibfnamefont{T.~S.}\ \bibnamefont{Chin}}, \bibinfo {author}
  {\bibfnamefont{T.~T.}\ \bibnamefont{Shun}}, \bibinfo {author}
  {\bibfnamefont{C.~H.}\ \bibnamefont{Tsau}},\ and\ \bibinfo {author}
  {\bibfnamefont{S.~Y.}\ \bibnamefont{Chang}},\ }%
  \bibfield{journal}{%
  \bibinfo {journal} {Adv. Eng. Mater.}\ }%
  \textbf{\bibinfo {volume} {6}},\ \bibinfo {pages} {299} (\bibinfo {year}
  {2004})%
  \bibAnnoteFile{NoStop}{Yeh04}%
\bibitem{Senkov10}%
  \BibitemOpen
  \bibfield{author}{%
  \bibinfo {author} {\bibfnamefont{O.~N.}\ \bibnamefont{Senkov}}, \bibinfo
  {author} {\bibfnamefont{G.~B.}\ \bibnamefont{Wilks}}, \bibinfo {author}
  {\bibfnamefont{D.~B.}\ \bibnamefont{Miracle}}, \bibinfo {author}
  {\bibfnamefont{C.~P.}\ \bibnamefont{Chuang}},\ and\ \bibinfo {author}
  {\bibfnamefont{P.~K.}\ \bibnamefont{Liaw}},\ }%
  \bibfield{journal}{%
  \bibinfo {journal} {Intermetallics}\ }%
  \textbf{\bibinfo {volume} {18}},\ \bibinfo {pages} {1758} (\bibinfo {year}
  {2010})%
  \bibAnnoteFile{NoStop}{Senkov10}%
\bibitem{Guo2011}%
  \BibitemOpen
  \bibfield{author}{%
  \bibinfo {author} {\bibfnamefont{S.}~\bibnamefont{Guo}}, \bibinfo {author}
  {\bibfnamefont{C.}~\bibnamefont{Ng}}, \bibinfo {author}
  {\bibfnamefont{J.}~\bibnamefont{Lu}},\ and\ \bibinfo {author}
  {\bibfnamefont{C.~T.}\ \bibnamefont{Liu}},\ }%
  \bibfield{journal}{%
  \bibinfo {journal} {J. Appl. Phys.}\ }%
  \textbf{\bibinfo {volume} {109}},\ \bibinfo {pages} {103505} (\bibinfo {year}
  {2011})%
  \bibAnnoteFile{NoStop}{Guo2011}%
\bibitem{Yang2012}%
  \BibitemOpen
  \bibfield{author}{%
  \bibinfo {author} {\bibfnamefont{X.}~\bibnamefont{Yang}}\ and\ \bibinfo
  {author} {\bibfnamefont{Y.}~\bibnamefont{Zhang}},\ }%
  \bibfield{journal}{%
  \bibinfo {journal} {Mat. Chem. and Phys.}\ }%
  \textbf{\bibinfo {volume} {132}},\ \bibinfo {pages} {233} (\bibinfo {year}
  {2012})%
  \bibAnnoteFile{NoStop}{Yang2012}%
\bibitem{Zhang2012}%
  \BibitemOpen
  \bibfield{author}{%
  \bibinfo {author} {\bibfnamefont{Y.}~\bibnamefont{Zhang}}, \bibinfo {author}
  {\bibfnamefont{X.}~\bibnamefont{Yang}},\ and\ \bibinfo {author}
  {\bibfnamefont{P.~K.}\ \bibnamefont{Liaw}},\ }%
  \bibfield{journal}{%
  \bibinfo {journal} {JOM}\ }%
  \textbf{\bibinfo {volume} {64}},\ \bibinfo {pages} {830} (\bibinfo {year}
  {2012})%
  \bibAnnoteFile{NoStop}{Zhang2012}%
\bibitem{Otto2013}%
  \BibitemOpen
  \bibfield{author}{%
  \bibinfo {author} {\bibfnamefont{F.}~\bibnamefont{Otto}}, \bibinfo {author}
  {\bibfnamefont{Y.}~\bibnamefont{Yang}}, \bibinfo {author}
  {\bibfnamefont{H.}~\bibnamefont{Bei}},\ and\ \bibinfo {author}
  {\bibfnamefont{E.~P.}\ \bibnamefont{George}},\ }%
  \bibfield{journal}{%
  \bibinfo {journal} {Acta Mater.}\ }%
  \textbf{\bibinfo {volume} {61}},\ \bibinfo {pages} {2628} (\bibinfo {year}
  {2013})%
  \bibAnnoteFile{NoStop}{Otto2013}%
\bibitem{Tian2013}%
  \BibitemOpen
  \bibfield{author}{%
  \bibinfo {author} {\bibfnamefont{F.~Y.}\ \bibnamefont{Tian}}, \bibinfo
  {author} {\bibfnamefont{L.~K.}\ \bibnamefont{Varga}}, \bibinfo {author}
  {\bibfnamefont{N.}~\bibnamefont{Chen}}, \bibinfo {author}
  {\bibfnamefont{L.}~\bibnamefont{Delczeg}},\ and\ \bibinfo {author}
  {\bibfnamefont{L.}~\bibnamefont{Vitos}},\ }%
  \bibfield{journal}{%
  \bibinfo {journal} {Phys. Rev. B}\ }%
  \textbf{\bibinfo {volume} {87}},\ \bibinfo {pages} {075144} (\bibinfo {year}
  {2013})%
  \bibAnnoteFile{NoStop}{Tian2013}%
\bibitem{Singh2011}%
  \BibitemOpen
  \bibfield{author}{%
  \bibinfo {author} {\bibfnamefont{S.}~\bibnamefont{Singh}}, \bibinfo {author}
  {\bibfnamefont{N.}~\bibnamefont{Wanderka}}, \bibinfo {author}
  {\bibfnamefont{B.~S.}\ \bibnamefont{Murty}}, \bibinfo {author}
  {\bibfnamefont{U.}~\bibnamefont{Glatzel}},\ and\ \bibinfo {author}
  {\bibfnamefont{J.}~\bibnamefont{Banhart}},\ }%
  \bibfield{journal}{%
  \bibinfo {journal} {Acta Mater.}\ }%
  \textbf{\bibinfo {volume} {59}},\ \bibinfo {pages} {182} (\bibinfo {year}
  {2011})%
  \bibAnnoteFile{NoStop}{Singh2011}%
\bibitem{Chen2006}%
  \BibitemOpen
  \bibfield{author}{%
  \bibinfo {author} {\bibfnamefont{M.~R.}\ \bibnamefont{Chen}}, \bibinfo
  {author} {\bibfnamefont{S.~J.}\ \bibnamefont{Lin}}, \bibinfo {author}
  {\bibfnamefont{J.~W.}\ \bibnamefont{Yeh}}, \bibinfo {author}
  {\bibfnamefont{S.~K.}\ \bibnamefont{Chen}}, \bibinfo {author}
  {\bibfnamefont{Y.~S.}\ \bibnamefont{Huang}},\ and\ \bibinfo {author}
  {\bibfnamefont{C.~P.}\ \bibnamefont{Tu}},\ }%
  \bibfield{journal}{%
  \bibinfo {journal} {Mat. Trans}\ }%
  \textbf{\bibinfo {volume} {47}},\ \bibinfo {pages} {1395} (\bibinfo {year}
  {2006})%
  \bibAnnoteFile{NoStop}{Chen2006}%
\bibitem{Zhang2013}%
  \BibitemOpen
  \bibfield{author}{%
  \bibinfo {author} {\bibfnamefont{Y.}~\bibnamefont{Zhang}}, \bibinfo {author}
  {\bibfnamefont{T.~T.}\ \bibnamefont{Zuo}}, \bibinfo {author}
  {\bibfnamefont{Y.~Q.}\ \bibnamefont{Cheng}},\ and\ \bibinfo {author}
  {\bibfnamefont{P.~K.}\ \bibnamefont{Liaw}},\ }%
  \bibfield{journal}{%
  \bibinfo {journal} {Sci. Rep.}\ }%
  \textbf{\bibinfo {volume} {3}},\ \bibinfo {pages} {1455} (\bibinfo {year}
  {2013})%
  \bibAnnoteFile{NoStop}{Zhang2013}%
\bibitem{Villars83}%
  \BibitemOpen
  \bibfield{author}{%
  \bibinfo {author} {\bibfnamefont{P.}~\bibnamefont{Villars}},\ }%
  \bibfield{journal}{%
  \bibinfo {journal} {J. Less Common Metals}\ }%
  \textbf{\bibinfo {volume} {92}},\ \bibinfo {pages} {215} (\bibinfo {year}
  {1983})%
  \bibAnnoteFile{NoStop}{Villars83}%
\bibitem{Pettifor88}%
  \BibitemOpen
  \bibfield{author}{%
  \bibinfo {author} {\bibfnamefont{D.~G.}\ \bibnamefont{Pettifor}},\ }%
  \bibfield{journal}{%
  \bibinfo {journal} {Materials Science and Technology}\ }%
  \textbf{\bibinfo {volume} {4}},\ \bibinfo {pages} {675} (\bibinfo {year}
  {1988})%
  \bibAnnoteFile{NoStop}{Pettifor88}%
\bibitem{Blum05}%
  \BibitemOpen
  \bibfield{author}{%
  \bibinfo {author} {\bibfnamefont{V.}~\bibnamefont{Blum}}\ and\ \bibinfo
  {author} {\bibfnamefont{A.}~\bibnamefont{Zunger}},\ }%
  \bibfield{journal}{%
  \bibinfo {journal} {Phys. Rev. B}\ }%
  \textbf{\bibinfo {volume} {72}},\ \bibinfo {pages} {02010R} (\bibinfo {year}
  {2005})%
  \bibAnnoteFile{NoStop}{Blum05}%
\bibitem{Curtarolo08}%
  \BibitemOpen
  \bibfield{author}{%
  \bibinfo {author} {\bibfnamefont{S.}~\bibnamefont{Curtarolo}}, \bibinfo
  {author} {\bibfnamefont{D.}~\bibnamefont{Morgan}},\ and\ \bibinfo {author}
  {\bibfnamefont{G.}~\bibnamefont{Ceder}},\ }%
  \bibfield{journal}{%
  \bibinfo {journal} {CALPHAD}\ }%
  \textbf{\bibinfo {volume} {29}},\ \bibinfo {pages} {163} (\bibinfo {month}
  {September}\ \bibinfo {year} {2005})%
  \bibAnnoteFile{NoStop}{Curtarolo08}%
\bibitem{Turchi01}%
  \BibitemOpen
  \bibfield{author}{%
  \bibinfo {author} {\bibfnamefont{P.~E.~A.}\ \bibnamefont{Turchi}}, \bibinfo
  {author} {\bibfnamefont{A.}~\bibnamefont{Gonis}}, \bibinfo {author}
  {\bibfnamefont{V.}~\bibnamefont{Drchal}},\ and\ \bibinfo {author}
  {\bibfnamefont{J.}~\bibnamefont{Kudrnovsk\'{y}}},\ }%
  \bibfield{journal}{%
  \bibinfo {journal} {Phys. Rev. B}\ }%
  \textbf{\bibinfo {volume} {64}},\ \bibinfo {pages} {085112} (\bibinfo {month}
  {August}\ \bibinfo {year} {2001})%
  \bibAnnoteFile{NoStop}{Turchi01}%
\bibitem{Turchi04}%
  \BibitemOpen
  \bibfield{author}{%
  \bibinfo {author} {\bibfnamefont{P.~E.~A.}\ \bibnamefont{Turchi}}, \bibinfo
  {author} {\bibfnamefont{V.}~\bibnamefont{Drchal}}, \bibinfo {author}
  {\bibfnamefont{J.}~\bibnamefont{Kudrnovsky}}, \bibinfo {author}
  {\bibfnamefont{C.}~\bibnamefont{Colinet}}, \bibinfo {author}
  {\bibfnamefont{L.}~\bibnamefont{Kaufman}},\ and\ \bibinfo {author}
  {\bibfnamefont{Z.-K.}\ \bibnamefont{Liu}},\ }%
  \bibfield{journal}{%
  \bibinfo {journal} {Phys. Rev. B.}\ }%
  \textbf{\bibinfo {volume} {71}},\ \bibinfo {pages} {094206} (\bibinfo {month}
  {March}\ \bibinfo {year} {2005})%
  \bibAnnoteFile{NoStop}{Turchi04}%
\bibitem{Hart05}%
  \BibitemOpen
  \bibfield{author}{%
  \bibinfo {author} {\bibfnamefont{G.~L.~W.}\ \bibnamefont{Hart}}, \bibinfo
  {author} {\bibfnamefont{V.}~\bibnamefont{Blum}}, \bibinfo {author}
  {\bibfnamefont{M.~J.}\ \bibnamefont{Malorski}},\ and\ \bibinfo {author}
  {\bibfnamefont{A.}~\bibnamefont{Zunger}},\ }%
  \bibfield{journal}{%
  \bibinfo {journal} {Nat. Mater.}\ }%
  \textbf{\bibinfo {volume} {4}},\ \bibinfo {pages} {391} (\bibinfo {year}
  {2005})%
  \bibAnnoteFile{NoStop}{Hart05}%
\bibitem{Kresse96}%
  \BibitemOpen
  \bibfield{author}{%
  \bibinfo {author} {\bibfnamefont{G.}~\bibnamefont{Kresse}}\ and\ \bibinfo
  {author} {\bibfnamefont{J.}~\bibnamefont{Furthmuller}},\ }%
  \bibfield{journal}{%
  \bibinfo {journal} {Phys. Rev. B}\ }%
  \textbf{\bibinfo {volume} {54}},\ \bibinfo {pages} {11169} (\bibinfo {year}
  {1996})%
  \bibAnnoteFile{NoStop}{Kresse96}%
\bibitem{Kresse99}%
  \BibitemOpen
  \bibfield{author}{%
  \bibinfo {author} {\bibfnamefont{G.}~\bibnamefont{Kresse}}\ and\ \bibinfo
  {author} {\bibfnamefont{D.}~\bibnamefont{Joubert}},\ }%
  \bibfield{journal}{%
  \bibinfo {journal} {Phys. Rev. B}\ }%
  \textbf{\bibinfo {volume} {59}},\ \bibinfo {pages} {1758} (\bibinfo {year}
  {1999})%
  \bibAnnoteFile{NoStop}{Kresse99}%
\bibitem{Perdew96}%
  \BibitemOpen
  \bibfield{author}{%
  \bibinfo {author} {\bibfnamefont{J.~P.}\ \bibnamefont{Perdew}}, \bibinfo
  {author} {\bibfnamefont{K.}~\bibnamefont{Burke}},\ and\ \bibinfo {author}
  {\bibfnamefont{M.}~\bibnamefont{Ernzerhof}},\ }%
  \bibfield{journal}{%
  \bibinfo {journal} {Phys. Rev. Lett.}\ }%
  \textbf{\bibinfo {volume} {77}},\ \bibinfo {pages} {3865} (\bibinfo {year}
  {1996})%
  \bibAnnoteFile{NoStop}{Perdew96}%
\bibitem{Neyts2013}%
  \BibitemOpen
  \bibfield{author}{%
  \bibinfo {author} {\bibfnamefont{E.~C.}\ \bibnamefont{Neyts}}\ and\ \bibinfo
  {author} {\bibfnamefont{A.}~\bibnamefont{Bogaerts}},\ }%
  \bibfield{journal}{%
  \bibinfo {journal} {Theor. Chem. Acc.}\ }%
  \textbf{\bibinfo {volume} {132}},\ \bibinfo {pages} {1320} (\bibinfo {year}
  {2013})%
  \bibAnnoteFile{NoStop}{Neyts2013}%
\bibitem{Grosso12}%
  \BibitemOpen
  \bibfield{author}{%
  \bibinfo {author} {\bibfnamefont{M.~F.}\ \bibnamefont{de~Grosso}}, \bibinfo
  {author} {\bibfnamefont{G.}~\bibnamefont{Bozzolo}},\ and\ \bibinfo {author}
  {\bibfnamefont{H.~O.}\ \bibnamefont{Mosca}},\ }%
  \bibfield{journal}{%
  \bibinfo {journal} {Physica B}\ }%
  \textbf{\bibinfo {volume} {407}},\ \bibinfo {pages} {3285} (\bibinfo {year}
  {2012})%
  \bibAnnoteFile{NoStop}{Grosso12}%
\bibitem{Bozzolo92}%
  \BibitemOpen
  \bibfield{author}{%
  \bibinfo {author} {\bibfnamefont{G.}~\bibnamefont{Bozzolo}}, \bibinfo
  {author} {\bibfnamefont{J.}~\bibnamefont{Ferrante}},\ and\ \bibinfo {author}
  {\bibfnamefont{J.~R.}\ \bibnamefont{Smith}},\ }%
  \bibfield{journal}{%
  \bibinfo {journal} {Phys. Rev. B}\ }%
  \textbf{\bibinfo {volume} {45}},\ \bibinfo {pages} {493} (\bibinfo {year}
  {1992})%
  \bibAnnoteFile{NoStop}{Bozzolo92}%
\bibitem{Jindo06}%
  \BibitemOpen
  \bibfield{author}{%
  \bibinfo {author} {\bibfnamefont{K.}~\bibnamefont{Masuda-Jindo}}, \bibinfo
  {author} {\bibfnamefont{V.~V.}\ \bibnamefont{Hung}},\ and\ \bibinfo {author}
  {\bibfnamefont{P.~E.~A.}\ \bibnamefont{Turchi}},\ }%
  \bibfield{journal}{%
  \bibinfo {journal} {Met. Mat. Trans. A}\ }%
  \textbf{\bibinfo {volume} {37}},\ \bibinfo {pages} {3403} (\bibinfo {year}
  {2006})%
  \bibAnnoteFile{NoStop}{Jindo06}%
\bibitem{Predmore70}%
  \BibitemOpen
  \bibfield{author}{%
  \bibinfo {author} {\bibfnamefont{R.}~\bibnamefont{Predmore}}\ and\ \bibinfo
  {author} {\bibfnamefont{R.~J.}\ \bibnamefont{Arsenault}},\ }%
  \bibfield{journal}{%
  \bibinfo {journal} {Scripta Met.}\ }%
  \textbf{\bibinfo {volume} {4}},\ \bibinfo {pages} {213} (\bibinfo {year}
  {1970})%
  \bibAnnoteFile{NoStop}{Predmore70}%
\end{thebibliography}%

\end{document}